**Insights from calculated phonon dispersion curves for an overlayer of H on Pt(111)**

Sampyo Hong [a], Talat S. Rahman [a,*], Rolf Heid [b], Klaus Peter Bohnen [b]

[a] *Department of Physics, Cardwell Hall, Kansas State University, Manhattan, KS 66506*

[b] *Forschungszentrum Karlsruhe, Institut fuer Festkoerperphysik, D-76021 Karlsruhe, Germany*

**Abstract**

We have calculated the dispersion curves of H vibrational modes on Pt(111), using first-principles, total energy calculations based on a mixed-basis set and norm-conserving pseudopotentials. Linear response theory and the harmonic approximation are invoked. For 1 ML coverage, H atoms are assumed to occupy the fcc hollow sites. At the Γ point of the surface Brillouin zone, we find modes, respectively, polarized parallel and perpendicular to the surface at 73.5 meV and 142.6 meV. The degeneracy of the parallel mode is lifted at the zone boundaries, yielding modes at 69.6 meV and 86.3 meV, at the M point, and at 79.4 meV and 80.8 meV, at the K point. The substrate surface modes are also found to shift in frequencies from their calculated values for clean Pt(111). We discuss the details of the changes in surface force constant on H adsorption on Pt(111). We also consider the case of subsurface adsorption for 2 ML of H and present vibrational frequencies of H atoms adsorbed in several subsurface sites. The appearance of new vertically polarized modes in the range of 98 – 106 meV (octahedral site) and 124 – 162 meV (tetrahedral site) is discussed in the context of experimental data.

1. Introduction

After extensive debate, three decades of experimental and theoretical studies of H on Pt(111) have led to the conclusion that H adsorbs preferably in the threefold hollow site, forms an ordered (1x1) overlayer at saturation coverage, and hybridization occurs between the 1s states of H and the d bands of Pt. The lingering issue is that of the assignment of the vibrational modes observed for full coverage which also raises questions about the symmetry of the adsorption site. While the very early high resolution electron energy loss spectroscopy (HREELS) measurements of Baro et al [1]. had revealed only two modes, one at 68 meV and another at 153 meV, at the Brillouin zone center, subsequent HREELS measurements have displayed an additional mode at about 112 meV [2,3]. The presence of the third mode has rekindled the debate on the preferred site for H adsorption [1,2,4–9], the polarization of the modes [1-3,10], and whether H remains localized or delocalized on the surface [3,11–14]. The issue of the adsorption site is all the more intriguing because in the thermal desorption data taken by Jacobi and coworkers there is the onset of an additional peak for H coverage exceeding an estimated 0.75 ML [3]. The diffusion barrier for H on Pt(111) has also been found to be small (~70 meV) [15], giving further testimony to the equivalence of the hollow sites: fcc and hcp. Related theoretical calculations assign the modes at 68 meV and 153 meV to vibrations localized around an equilibrium position in the three fold hollow site [3]. The mode at 112 meV is not easily accounted for in these calculations, for obvious reasons. The symmetry of the surface requires that the two in-plane modes be degenerate at the Brillouin zone center, if H occupies a single adsorption site. The third modes has thus been explained as either an overtone, a combination loss [2], or as a mixed mode arising from H delocalized over the fcc and hcp hollow sites [3]. On the other hand, a more complex adsorption scheme could lead to some interesting conclusions about H vibrational frequencies. To explain details in the analysis of photoemission data, a two-state adsorption model was proposed [4,16], but subsequently argued to be incompatible with photoemission data [7]. The proposal of subsurface H has reemerged in recent times [14,17], although clear experimental evidence is yet to be provided.

In this study we consider the case of 1 ML of H adsorbed in the fcc-hollow sites on Pt(111). For a comparative study of the vibrational dynamics of the system for the three adsorption sites: fcc-hollow, hcp-hollow, and on-top is given in Ref. 18. Our calculations are based on density functional perturbation theory (DFPT) based on the linear response theory and the harmonic approximation [19]. Since proposal of subsurface H has reemerged in recent times, we consider also the case of subsurface H adsorption for 2 ML coverage. This higher coverage is not unreasonable since the exact knowledge of coverage from experiments performed for saturation coverage is not easy to obtain, particularly for an adsorbate like H.

Below we provide some details of our calculations in Section II, followed by a description of our model system in section III. The results are presented in Section IV with discussion of the calculated structural properties followed by that of the phonon dispersion curves. The conclusion are summarized in Section V.

2. **Theoretical details**

The total energy of the fully relaxed systems was calculated using the density functional theory within the pseudopotential method in a mixed basis representation [20]. To represent ion-electron interaction, a norm-conserving pseudopotential was used in the local density approximation (LDA). The electron-electron interaction [21] was represented by a Hedin-Lundqvist form of the exchange-correlation functional [22]. For the valence states of Pt and H local functions of d and s type with radial cutoffs of 2.1 a.u. and 0.7 a.u., respectively, were applied. Kinetic energy cutoff for plane waves was 16.5 Ryd. Integration over an irreducible Brillouin-zone were carried out using 42 special kpoints. A Fermi level smearing of 0.014 Ryd was employed [23]. Structural relaxations were carried out until the forces, calculated using the Hellmann-Feynman theorem, on all atoms were less than 0.001 Ryd/a.u.

The calculations of the vibrational frequencies and their eigenvector were performed using linear response theory within a perturbative approach in a mixed-basis representation of the wave functions [24]. Phonon dispersion curves were obtained by standard Fourier interpolation method using a (6x6) q-point mesh. Surface force constants were merged with bulk force constants and an asymmetric bulk slab of 100 layers was added to obtain projected bulk phonon modes [25] using standard lattice dynamical methods. An (8x8x8) q-point mesh was used for the calculation of the bulk phonons. Vibrational modes were labeled as surface modes if their eigenvectors contained contributions larger than 20% of the displacements of the atoms in the top two Pt layers.

In Fig. 1(a) the top view of a close-packed arrangement of spheres representing a segment of the fcc(111) surface is shown. The large white circles represent Pt atoms in the first layer, while the small black circles denote the fcc-hollow sites, and the dotted circles represent Pt atoms in second layer. The two dimensional Brillouin zone for fcc(111) surface is shown in Fig. 1(b). For the calculations a total of 9 layers of Pt atoms and one H atom on each side were used for a slab with inversion symmetry. The surface (1x1) unit-cell used to simulate 1ML and 2ML H adsorption is shown in Fig. 1(a) as a parallelogram. The side view of the surface system presented in Fig. 1c relates to sub-surface adsorption which we discuss later in section 3.

## 3. RESULTS AND DISCUSSIONS

### 3.1. Structural Relaxations

We find the height of the H atoms above the Pf surface to be 0.93 Å. The corresponding H-Pt bond length is 1.85 Å and the adsorption energy is calculated to be 0.809 eV. These results are in agreement with previous LDA results [8] which, when compared to experimental value (0.47 eV) is an overestimation. On the other hand, we find that H causes an outward relaxation of the Pt surface layer of 2.7% (fcc) [26], in agreement with experimental observations.

### 3.2. Surface phonon dispersion curves

The surface phonon dispersion curves for 1ML of H on Pt(111) calculated along two high symmetry directions, $\Gamma - M$ and $\Gamma - K$, in the surface Brillouin zone are shown in Fig. 2. The grey solid lines correspond to bulk-projected modes, the dark filled-circles are the Pt substrate surface modes in the lower panel and the vibrations of H atoms in the upper panel, and the white circles represent the data from He-atom surface (HAS) measurements [27]. It should be noted that an asymmetric slab filing was performed which gave two types of surfaces in the model system: one H-covered and the other Pt bulk-terminated. Each of these surfaces contributed one set of curves well separated from the bulk-projected band. The grey line below the bulk-projected band, in Fig. 2 is the one originating from the bulk-terminated Pt surface and should be ignored as it is an artifact of our calculational set up. The first lowest energy surface mode in Figs. 2(a), at the zone boundaries, is the Rayleigh Wave (RW) which is easily identifiable since it is generally well separated from the bulk-projected bands and usually has a vertical polarization at the Brillouin zone boundary.

At the M point, the dispersion curves display three Pt substrate surface modes with frequencies 9.6 meV, 21.9 meV, and 22.3 meV. The RW (9.6 meV) consists of purely vertical vibrations of the atoms in the top Pt layer. The next two modes (21.9 meV, and 22.3 meV) comprise purely longitudinal displacements of the atoms in the first and second Pt layers. At the K point, we find four surface modes with frequencies 9.7 meV, 13.3 meV, 17.4 meV, and 19.6 meV. The frequency of the RW (9.7 meV) is in good agreement with that in the HAS measurements [27]. The second substrate surface mode at K (13.3 meV) is mainly of vertical character with a weak parallel component. The three topmost Pt layers take part in this mode such that there is a vertical displacement of Pt atoms in the second layer while the atoms in the first and third Pt layers vibrate in the direction parallel to the surface. The third substrate mode (17.4 meV) is polarized purely in the surface plane, however, its amplitude extends below to five Pt layers. The fourth mode (19.6 meV) is a surface resonance and its amplitude penetrates deep into the bulk. The dispersions of the H vibrational modes are represented by the top three curves in Fig. 2. The two lower curves arise from in-plane vibrations, while the top curve is the result of vertical displacements of the H atom. The parallel vibrational modes show substantial dispersion with a maximum split of 16.6 meV at the M point. The upper branch consists of longitudinal modes, while the lower one is the shear-horizontal mode, along both the $\Gamma$- M and $\Gamma$- K directions. At the K point, the two H parallel modes are almost degenerate with a maximum difference in frequency of only 1.4 meV.

It is interesting to quantify the impact of H adsorption on Pt surface phonon dispersions, since it is so often assumed that the substrate phonons remain unchanged in the presence of an adsorbate as light as H [27]. To get a better estimate of the shifts in the frequencies on H adsorption, the calculated surface phonon dispersion curves for clean Pt(111) are displayed together with experimental values (white circles) in Fig. 2. Detailed analysis of the results in Fig. 2 can be found in a recent review article [28]. We find that the surface modes for clean Pt(111) undergo a few changes on H adsorption: two RW or RW-like modes on the clean surface, near the zone boundaries and along the M – K direction, are replaced by only one in the dispersion curves for the H-covered surface. This is because the first mode which is vertically polarized (RW), at the zone boundaries, is softened and the next two surface modes (polarized in the surface plane) are stiffened such that they are pushed above into the bulk hand, on H adsorption in the fcc-site (Table II). Clearly H adsorption alters the force field of the surface atoms and consequently the surface phonon dispersion curves. Below we discuss some of the highlights of the Pt(111) surface force constant changes brought about on H adsorption.

### 3.3. Force constant changes

The calculated force constant matrix for the Pt(111) surface atoms with 1 ML coverage of H exhibit substantial changes from the values on the clean Pt(111) surface. To fully appreciate the changes in the strength of coupling between any two atoms we need to take into account all components in the force-constant matrix $\Phi_{\alpha\beta}(\sigma)$ for a bond $\sigma$. For this purpose, the convenient concept of an average coupling strength $<I>$ for a bond $\sigma$ is introduced [28]. It is defined as

$$<I> \equiv \sqrt{\frac{1}{3}\sum_{\alpha\beta}\Phi^2_{\alpha\beta}(\sigma;i,j)}$$

where $\alpha$ and $\beta$ are Cartesian coordinates, and $\sigma$ is the bond length between atoms i and j. In Table I and II values of $\Phi_{\alpha\beta}$ between Pt atoms in the surface layers are presented together with those of atoms on clean Pt(111), and in bulk Pt. In Table I, the first three columns display the force constant components between two Pt atoms in the topmost layer separated by first nearest neighbor distance (1NN). Similarly the next three columns correspond to force constants between second nearest neighbors (2NN). For 1NN interaction, the average coupling strength $<I>$ between top layer Pt atoms, in the presence of H in fcc hollow sites, is 42658.6 dyn/cm. This is to be compared with force constants of 22675.6 dyn/cm, and 33633.3 dyn/cm for Pt atoms on clean Pt(111) and in the bulk, respectively. For 2NN interaction, the average force constant reduces to 799.2 dyn/cm for Pt atoms with a H overlayer, to 3438.7 dyn/cm for clean Pt(111), and to 1603.8 dyn/cm for Pt atoms in the bulk. Thus, for fcc site adsorptions the force constants for lateral interaction between Pt atoms in the top layer undergo a stiffening of about 26% from bulk. The stiffening of these force constants leads to shift towards higher frequencies of the surface mode in the stomach gap, as can be seen by the shift of the mode to 21.9 meV in the dispersion curves in Figs. 2(a). This is understandable since this stomach gap mode consists mainly of in-plane motion of the Pt surface atoms. Consistent with previous results [27], we find a remarkable weakening (33%) of the lateral interaction between Pt atoms in the top layer on the clean surface from the bulk value. The force constants for interlayer coupling are presented in Table II. As in Table I, the first three columns contain force constant components corresponding to first nearest neighbor bond connecting Pt atoms in the topmost layer to those in the second layer, while the next three columns display the ones for first nearest bond of H-Pt. The interlayer average coupling strength <I> for fcc, clean, and bulk are 27788.1 dyn/cm, 33083.5 dyn/cm, and 33583.8 dyn/cm, respectively. Interestingly the interlayer average-coupling strength hardly changes for the clean Pt(111) surface with respect to bulk (even 3% softening) but it softens by 18% for fcc adsorption in perfect agreement with observed structural relaxations of the topmost layer [28]. Figure 3 shows change of in-plane [Fig. 3(a)] and interlayer [Fig. 3(b)] average force constants with respect to nearest neighbor distances. Moderate description of lattice dynamics may be obtained by taking into account at least up to $4^{th}$ NN, but for best up to $6^{th}$ NN force constants may be necessary [27].

### 3.4. H vibrational modes calculated for 2 ML coverage

As is well known, the exact determination of coverage in experiments is non-trivial. In the experimental measurements of H vibrational frequencies by Wang et al. [3], it has been suggested that the saturation coverage may correspond to more than 1 ML which begs the question of where the other monolayer goes. One possibility is the occupation of two types of sites on the surface, i.e. 1 ML in either the hcp hollows or the top sites, in addition to the 1 ML on fcc hollows. As mentioned in Introduction, the possibility of subsurface adsorption has also been the subject of interrogation in several recent publications. Present calculations show that subsurface adsorption is not likely for less than 1ML coverage and H atoms need to overcome energy barriers of about 0.67 eV and 1.53 eV , respectively, for 1.25 ML and for 2 ML coverage, to reach the subsurface region. It is unlikely that under conditions of low temperature and pressure H atoms will overcome such barriers. However, it has been argued that at 500K 1 bar of $H_2$ is sufficient to introduce 0.25 ML of H in the subsurface while maintaining 1 ML on the surface[17].

Subsurface adsorption at high temperature and pressure is thus a possibility, which remains to be confirmed by experiments. With this in mind, we have carried out lattice dynamical calculations for configurations in which 1 ML remains on the surface (fcc-site) and 1 ML is absorbed between the first and second Pt layers. Our calculated frequencies of H vibrational modes, at the Γ point of the two dimensional Brillion zone, for such configurations are presented in Table III along with binding energies per H atom and the location of H atoms from the topmost Pt layer. In the Table $d_{H1-Pt1}$, $d_{H2-Pt1}$ are vertical heights of two hydrogens from the topmost Pt layer and accordingly negative height corresponds to subsurface adsorption. The last column in the Table III contains frequencies of H atoms at Γ point and their polarizations in parenthesis where underlined numbers are frequencies of subsurface H atoms, and capital letter P and V represent parallel and vertical polarizations, respectively, and the letters f, h, and t denote H adsorption at fcc, hcp, and on-top adsorption sites on the surface, respectively. Binding energies were calculated using the following equation: $E_{binding}$[H atom] = ( E[H/Pt(111)] – E[clean Pt(111)] – 2*E[isolated $H_2$ molecule] )/2 as we use symmetrical slab with 2ML H on both sides.

Before stressing the validity of sub-surface adsorption when H coverage exceeds 1 ML, we should mention that we have also considered the case when the two monolayers are on the surface (row 5 and 6 in Table III). The only stable structure in this situation was found to be the simultaneous occupation of the fcc and on-top sites with binding energy of 0.38 eV per H atom which is one-half of that for the 1ML case (row 1 and 3). The frequencies of H parallel vibration for this two-site adsorption model were found to be 71 meV and 96 meV and those for vibration normal to the surface were 132 meV and 259 meV. Note that this two-site adsorption model fails to of predict the wanted mode (106 meV with vertical polarization) and thus does not present itself as the possible candidate. Let us then return to the discussion of sub-surface adsorption. Assuming that H atoms adsorbed at fcc sites in the surface overcome the energy barrier so that the subsurface monolayer is formed in the octahedral position below the topmost Pt layer, there are two possible configurations of H atoms adsorbed at the top sites in the surface, i.e. they can diffuse either to fcc site or to hcp site ( row 6 and 7 in the Table III). Our calculations predict equivalency between the two configurations as shown in binding energies and vibrational frequencies of H atoms producing 51 meV and 74 meV for parallel vibration and two spectrums of vertical vibration: 98 - 106 meV and 148 meV. The former vertical modes comes from subsurface H atoms and the latter from surface H atoms. Starting with this configuration, it may be possible for the subsurface H atoms in the octahedral sites to diffuse to neighboring subsurface empty sites, leading to two further configurations: occupancy directly below the fcc site in the surface (labeled "tetrahedral a" in the Table III) or directly below Pt atoms in the surface (labeled "tetrahedral b" in the Table III) (see also Fig. 1c). These two configurations are shown in the last two rows in the Table III. They have the largest binding energy among all configurations considered and equivalent to each other while their vibration frequencies differ depending on vertical vibrations of the subsurface H atoms. For instance, the vertical frequency of the subsurface H atoms for tetrahedral a is 162 meV while it is 124 meV for tetrahedral b, which are the reflection of subsurface geometry. Additionally modes at 46 meV and 98 meV for parallel vibration, and at 124 meV and 148 – 162 meV for vertical vibration are produced. As compared to frequencies of H vibrations for the case of 1 ML coverage, the most important change is the appearance of new vertical polarized modes in the range of either 98 – 106 meV (octahedral site) or 124 – 162 meV (tetrahedral site) which are all contributed from the subsurface H atoms. Note that for all configurations considered, vibrational frequencies of H atoms in the surface are hardly affected by subsurface adsorption.

One thing to note is that our calculated binding energies based on the local density approximation overestimates that obtained using the generalized gradient approximation (GGA) [8, 17] which is in closer agreement with the experimental data. However, we have consistently used LDA in our calculation for all results reported in this study and their relative effect should be reliable.

## 4. CONCLUSION

We have calculated the phonon dispersion curves for H on Pt(111), using first-principles, total energy calculations based on a mixed-basis set and norm-conserving pseudopotentials. Linear response theory and the harmonic approximation are invoked. At 1 ML coverage of H on Pt(111), we find the fcc adsorption at a height of 0.9 Å, with a binding energy of 0.809 eV. We find that H adsorption alters the force field of the surface in such way that the lateral interaction between the Pt atoms on the H-covered surface undergoes stiffening rather than softening exhibited by Pt surface atoms on the clean Pt(111). As a result, vertically polarized surface modes are lowered in energy while those polarized parallel to the surface are raised in energy. For instance, the frequency of the Pt Rayleigh wave at the zone boundaries is reduced and the Pt surface mode in the stomach gap disappears into the bulk band. For H vibrations we find modes polarized parallel and perpendicular to the surface at 73.5 meV and 142.6 meV at the $\Gamma$ point of the surface Brillouin zone, respectively. We carried out calculations for several configurations based on two-adsorption-site model consisting of H occupations of both sites such as either fcc and top, or fcc and hcp sites in the surface but failed to produce the vertically polarized mode (112 meV). As subsurface adsorption of H atoms for Pt(111) is a possibility, we present results for subsurface H adsorption for 2 ML coverage where 1 ML subsurface coverage is assumed to occur in octahedral or tetrahedral sites just below the topmost Pt layer while 1 ML remains in the surface. For octahedral subsurface adsorption, calculated frequencies range from 51 to 74 meV for parallel vibration, from 98 to 106 meV and 148 meV for vertical vibration. For tetrahedral subsurface adsorption, former ones were 46 meV and 87 meV, and latter ones were 124 meV and 145 – 162 meV. Vibrations of H atoms in surface are hardly affected by subsurface. We await further experiments particularly on H vibrations of the H-Pt(111) system to provide further answers to the issue of H on Pt(111).

**ACKNOWLEDGEMENTS**

The work was supported in part by the US National Science Foundation, Grant CHE-0205064. Computations were performed on the multiprocessors at NCSA, Urbana, and at Forschungszentrum, Karlsruhe. TSR also acknowledges the support of the Alexander von Humboldt Foundation and thanks her colleagues at the Fritz Haber Institut, Berlin and at the Forschungszentrum, Karlsruhe for their warm hospitality. We are grateful to Stefan Badescu and Karl Jacobi for getting us interested in the subject and for helpful discussions.

———————————————————————

[*]Corresponding author. E-mail address: rahman@phys.ksu.edu; WWW homepage: http://www.phys.ksu.edu/~rahman/; FAX: 785 532 6806.

**Figures**

FIG. 1. (a) the top view of the fcc(111) surface in which the large gray circles are Pt atoms in the first layer and the dotted circles represent Pt atoms in the second layer while the points labeled F and H are the fcc and hcp hollow sites, respectively. The top site for hydrogen adsorption is labeled T. The surface (1x1) unit-cell is shown as a parallelogram. (b) the side view of adsorption sites together with Pt layers in which O labels the octahedral sites ( the one just below the fcc hollow), and Ta and Tb indicates the two tetrahedral sites. (c) the surface Brillouin zone for the fcc(111) crystal.

FIG. 2. Surface phonon dispersion curves for (a) a monolayer H on Pt(111) (b) clean Pt(111).

FIG. 3. Average force constant changes for (a) Pt atoms in the topmost layer and (b) Pt atoms in the first and second layers versus nearest neighbor distances.

**Tables**

Table I. In-plane force constant matrix $\Phi_{\alpha\beta}$ for Pt atoms in the topmost layer compared with those for clean Pt(111) and for those in the bulk. The unit is dyn/cm.

| | | $\Phi_{\alpha\beta}$(1 NN) | | | $\Phi_{\alpha\beta}$(2 NN) | | |
|---|---|---|---|---|---|---|---|
| | | X | Y | Z | X | Y | Z |
| FCC | X | 52845.3 | 40276.1 | 5258.3 | -6.2 | -0.0 | -1230.2 |
| | Y | 26204.0 | 14463.0 | -3605.6 | -0.0 | -53.3 | 0.0 |
| | Z | 493.3 | -6356.7 | -8205.9 | 525.7 | 0.0 | 351.8 |
| CLEAN | X | 27053.9 | 17367.6 | -7719.6 | 5878.5 | 0.0 | -255.6 |
| | Y | 16326.4 | 7600.7 | -5641.1 | -0.0 | 765.6 | 0.0 |
| | Z | 8745.1 | 3864.8 | 1359.7 | -390.8 | 0.0 | -337.0 |
| BULK | X | 42154.0 | 27271.9 | 566.8 | 2506.2 | 0.0 | -196.4 |
| | Y | 27271.9 | 10663.1 | -981.8 | - 0.0 | -1141.3 | 0.0 |
| | Z | 566.8 | -981.8 | -3584.2 | -196.4 | 0.0 | -237.4 |

Table II. Interlayer force constant matrix $\Phi_{\alpha\beta}$ for Pt atoms in the topmost layer compared with those for clean Pt(111) and for those in the bulk. The unit is dyn/cm.

|  |  | $\Phi_{\alpha\beta}$(1 Pt layer, 2 Pt layer) | | | $\Phi_{\alpha\beta}$(1 Pt layer, H overlayer) | | |
|---|---|---|---|---|---|---|---|
|  |  | X | Y | Z | X | Y | Z |
| FCC | X | -14687.0 | 0.0 | 23355.1 | -322.9 | -5848.9 | -1457.0 |
| FCC | Y | 0.0 | 3848.1 | 0.0 | -5848.9 | -7076.5 | -2523.5 |
| FCC | Z | 28513.3 | 0.0 | -26973.1 | -19231.9 | -33310.6 | -24303.4 |
| CLEAN | X | -12405.0 | 0.0 | 29107.0 | | | |
| CLEAN | Y | 0.0 | 5891.2 | 0.0 | | | |
| CLEAN | Z | 30864.6 | 0.0 | -35987.7 | | | |
| BULK | X | -15641.8 | 0.0 | 29709.4 | | | |
| BULK | Y | 0.0 | 3627.5 | 0.0 | | | |
| BULK | Z | 30406.4 | 0.0 | -36312.4 | | | |

Table III. Relaxations, binding energies, and vibrational frequencies of H atoms for H/Pt(111). Here all the frequencies are calculated for Brillouin zone center and their polarization is given in parenthesis where underlined numbers are frequencies of subsurface H atoms, and capital letter P and V represent parallel and vertical polarizations, respectively, and small letter f, h, and t mean that H atom vibrates at fcc, hcp, and on-top adsorption sites on the surface, respectively.

| H coverage | H surface site | H subsurface site | Binding energy (eV) | $d_{H1-Pt1}, d_{H2-Pt1}$ (Å) | H frequency (meV) |
|---|---|---|---|---|---|
| 1 | fcc | | 0.809 | 0.93 | 73.5(Pf), 142.6(Vf) |
| 1 | hcp | | 0.763 | 0.93 | 67.3(Pf), 143.9(Vf) |
| 1 | top | | 0.706 | 1.56 | 47.4(Pf), 277.2(Vf) |
| 2 | fcc, hcp | | 0.014 | 0.97, 0.97 | Lattice dynamically unstable |
| 2 | fcc, top | | 0.383 | 0.82, 1.58 | 71(Pt), 96(Pf), 132(Vf), 259(Vt) |
| 2 | fcc | octahedral | 0.390 | 0.93, -1.83 | <u>51</u>(P), 72(Pf), <u>106</u>(V), 148(Vf) |
| 2 | hcp | octahedral | 0.383 | 0.92, -1.79 | <u>71</u>(P), 74(Ph), <u>98</u>(V), 148(Vh) |
| 2 | fcc | tetrahedral a | 0.446 | 0.93, -1.28 | <u>46</u>(P), 78(Pf), 148(Vf), <u>162</u>(V) |
| 2 | fcc | tetrahedral b | 0.459 | 0.89, -1.86 | <u>62</u>(P), 87(Pf), <u>124</u>(V), 145(Vf) |

1(a)

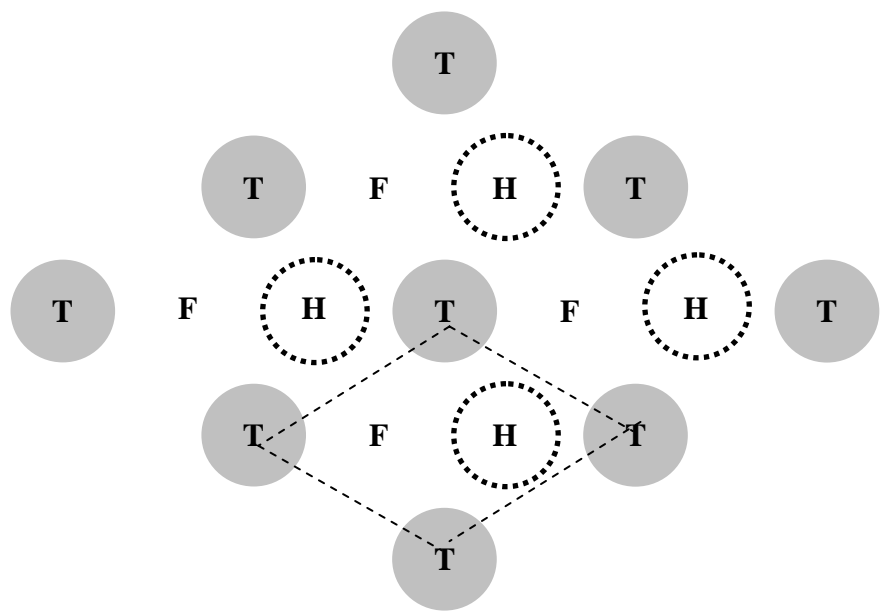

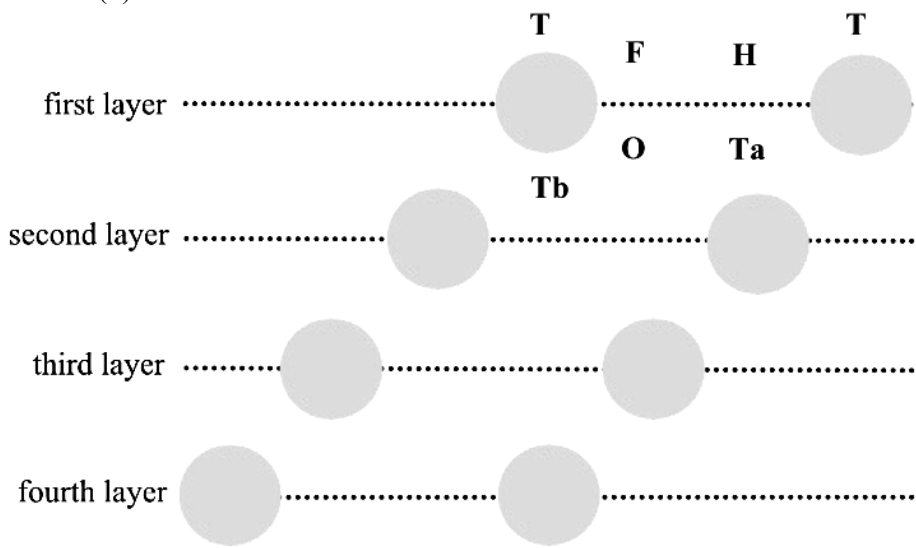
1(b)

1(c)

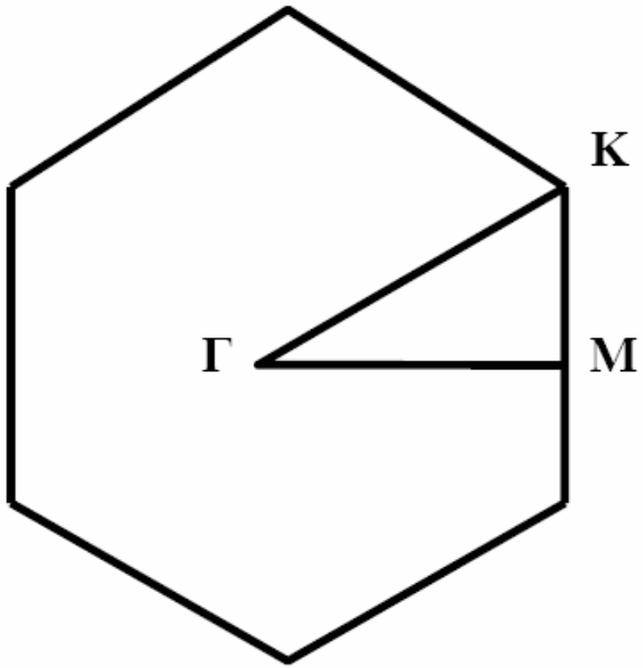

2(a)

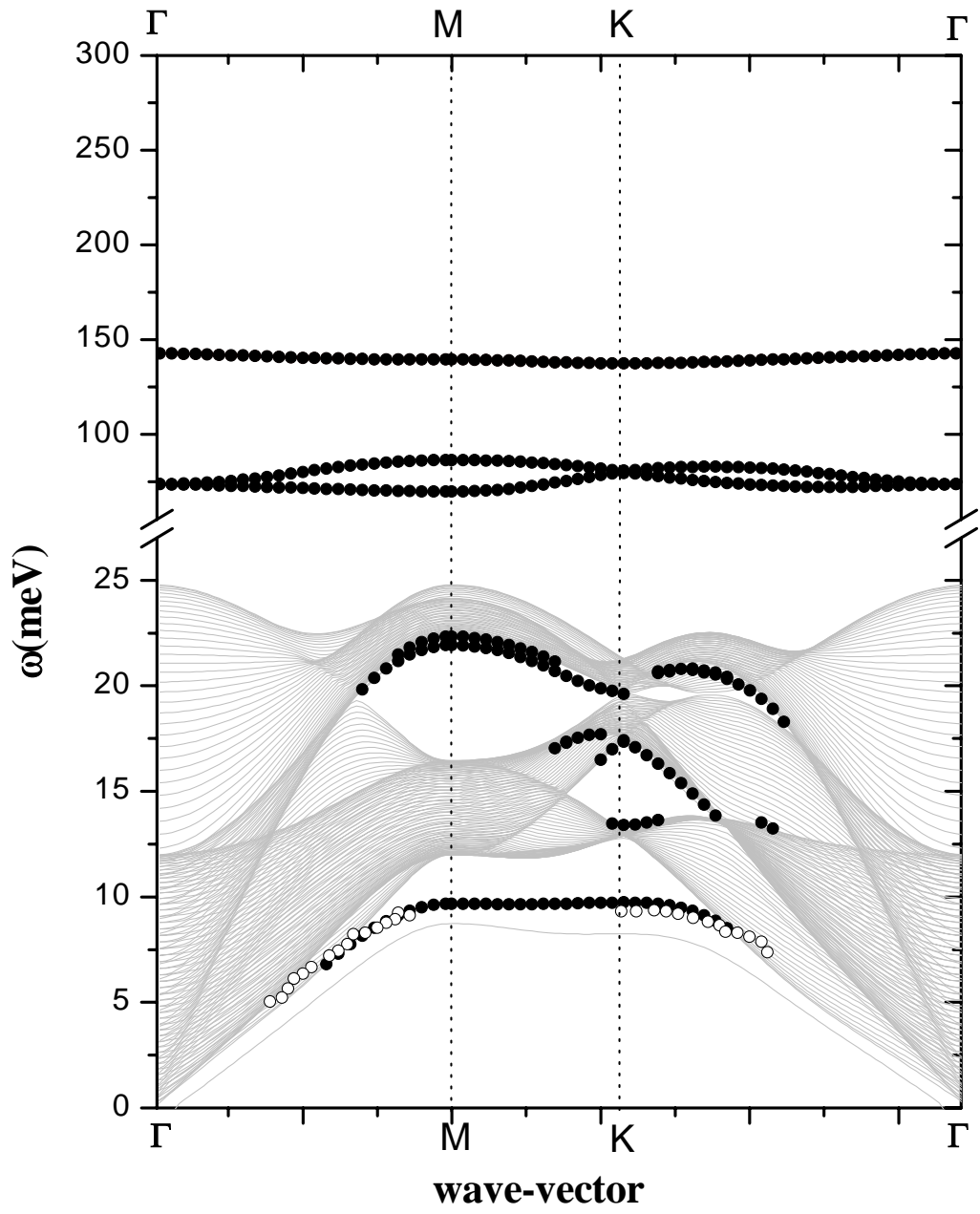

2(b)

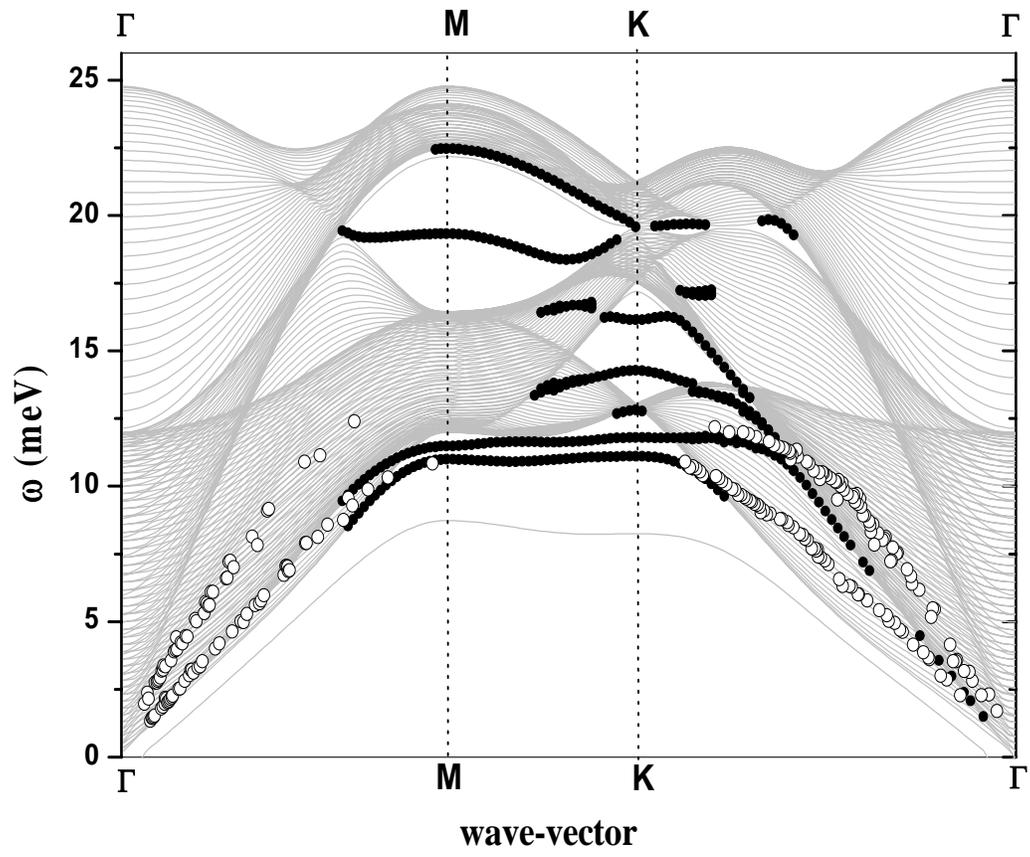

3(a)

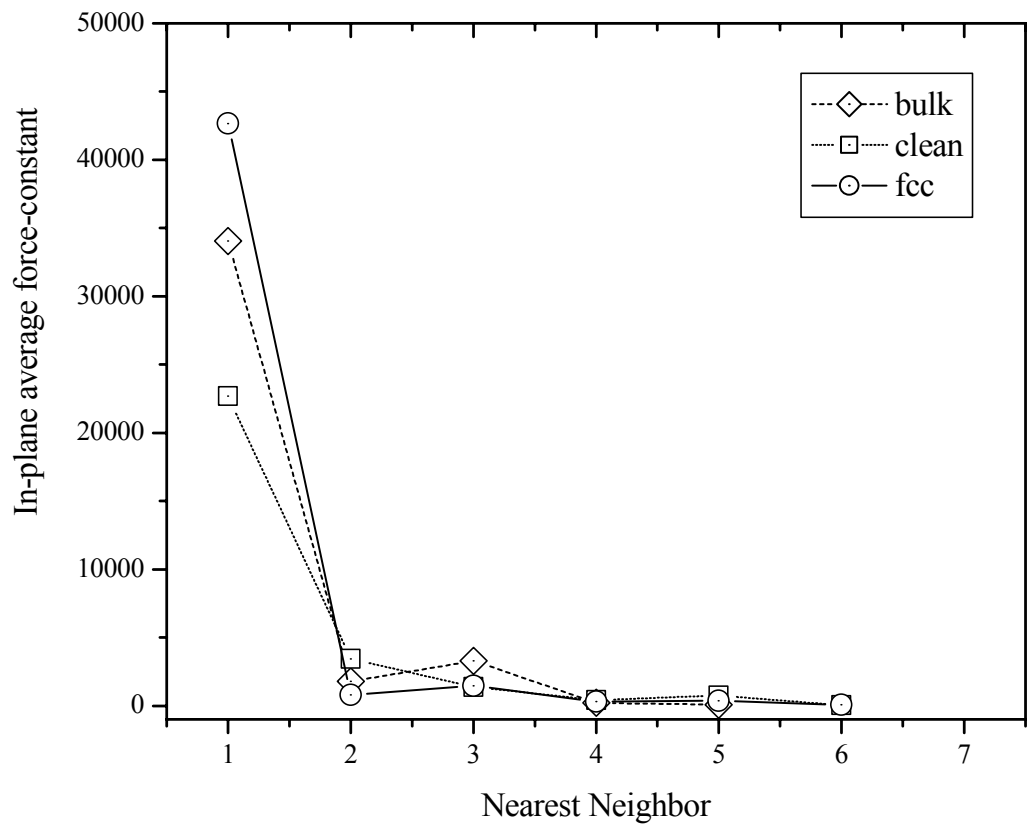

3(b)

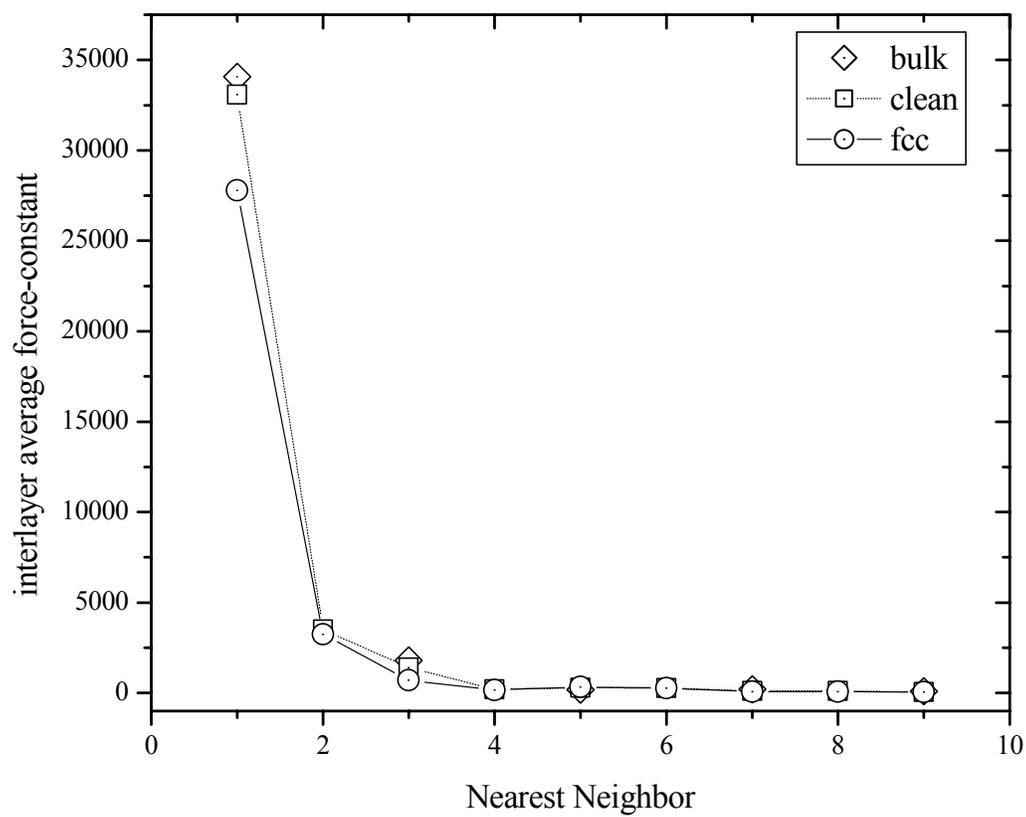